\documentclass[runningheads,anonymous]{llncs}

\usepackage{amsmath,amssymb,amsfonts}
\usepackage{marvosym}

\usepackage[colorlinks = true,
            linkcolor = blue,
            urlcolor  = blue,
            citecolor = blue,
            anchorcolor = blue]{hyperref}
\usepackage{graphicx}
\usepackage{array,multirow,graphicx}
\usepackage{float}
\usepackage{algorithmic}
\usepackage{graphicx}
\usepackage{textcomp}
\usepackage{xcolor}
\usepackage{hyperref}
\usepackage{cleveref}
\usepackage{footmisc}
\usepackage{xcolor,colortbl}
\usepackage{verbatim}
\usepackage{cite}

\definecolor{Gray}{gray}{0.85}
\definecolor{LightCyan}{rgb}{0.88,1,1}
\newcolumntype{a}{>{\columncolor{Gray}}c}
\newcolumntype{b}{>{\columncolor{white}}c}

\begin{document}
\title{VinDr-SpineXR: A deep learning framework for spinal lesions detection and classification from radiographs}
\titlerunning{A deep learning-based spinal lesions detection framework on X-rays}

\author{Hieu T. Nguyen\inst{1,2} \and
Hieu H. Pham\inst{1,3} \and
Nghia T. Nguyen\inst{1} \and
Ha Q. Nguyen\inst{1,3} \and
Thang Q. Huynh\inst{2} \and
Minh Dao\inst{1} \and
Van Vu\inst{1,4}}

\institute{Medical Imaging Center, Vingroup Big Data Institute, Hanoi, Vietnam \and
School of Information and Communication Technology, Hanoi University of Science and Technology, Hanoi, Vietnam  \and
College of Engineering \& Computer Science, VinUniversity, Hanoi, Vietnam \and
Department of Mathematics, Yale University, New Heaven, USA \\
All correspondence should be addressed to Hieu H. Pham at \email{v.hieuph4@vinbigdata.org}}

\maketitle
\begin{abstract}
Radiographs are used as the most important imaging tool for identifying spine anomalies in clinical practice. The evaluation of spinal bone lesions, however, is a challenging task for radiologists. This work aims at developing and evaluating a deep learning-based framework, named VinDr-SpineXR, for the classification and localization of abnormalities from spine X-rays. First, we build a large dataset, comprising 10,468 spine X-ray images from 5,000 studies, each of which is manually annotated by an experienced radiologist with bounding boxes around abnormal findings in 13 categories. Using this dataset, we then train a deep learning classifier to determine whether a spine scan is abnormal and a detector to localize 7 crucial findings amongst the total 13. The VinDr-SpineXR is evaluated on a test set of 2,078 images from 1,000 studies, which is kept separate from the training set. It demonstrates an area under the receiver operating characteristic curve (AUROC) of 88.61\% (95\% CI 87.19\%, 90.02\%) for the image-level classification task and a mean average precision (mAP@0.5) of 33.56\% for the lesion-level localization task. These results serve as a proof of concept and set a baseline for future research in this direction. To encourage advances, the dataset, codes, and trained deep learning models are made publicly available.

\end{abstract}

\section{Introduction}

\subsection{Spine X-ray interpretation}
Conventional radiography or X-ray has the ability to offer valuable information than many  other imaging modalities in the assessment of spinal lesions~\cite{priolo1998current,doi:10.1177/1755738013491081}. It has been the primary tool widely used to identify and monitor various abnormalities of the spine. A wide range of spine conditions can be observed from the X-rays like fractures, osteophytes, thinning of the bones, vertebral collapse, or tumors~\cite{baert2007spinal,rodallec2008diagnostic}. In clinical practice, radiologists usually interpret and evaluate the spine on X-ray scans stored in Digital Imaging and Communications in Medicine (DICOM) standard. Abnormal findings could be identified based on differences in density, intensity, and geometry of the lesions in comparison with normal areas. In many cases, those differences might be subtle and it requires an in-depth understanding of diagnostic radiography to spot them out. Large variability in the number, size, and general appearance of spine lesions makes the interpretation of spinal X-rays a complex and time-consuming task. These factors could lead to the risk of missing significant findings~\cite{pinto2018traumatic}, resulting in serious consequences for patients and clinicians.

The rapid advances in machine learning, especially deep neural networks, have demonstrated great potential in identifying diseases from medical imaging data~\cite{shen2017deep}. The integration of such systems in daily clinical routines could lead to a much more efficient, accurate diagnosis and treatment~\cite{krupinski2010long}. In this study, we aim to develop and validate a deep learning-based computer-aided diagnosis (CAD) framework called VinDr-SpineXR that is able to classify and localize abnormal findings from spine X-rays. Both the training and validation of our proposed system are performed on our own dataset where radiologists' annotations serve as a strong ground truth.

\subsection{Related work}
Developing CAD tools with a high clinical value to support radiologists in interpreting musculoskeletal (MSK) X-ray has been intensively studied in recent years~\cite{gundry2018computer}. Early approaches to the analysis of spine X-rays focus on using radiographic textures for the detection of several specific pathologies such as vertebral fractures~\cite{kasai2006development}, osteoporosis~\cite{kasai2006computerized}, or osteolysis~\cite{wilkie2007imputation}. Currently, deep convolutional networks~\cite{lecun2015deep} (CNNs) have shown their significant improvements for the MSK analysis from X-rays~\cite{mandal2015developing,sa2017intervertebral,lindsey2018deep,kim2020automatic,thian2019convolutional,zhang2020new,rajpurkar2017mura}. Most of these studies focus on automated fracture detection and localization~\cite{lindsey2018deep,kim2020automatic,thian2019convolutional,zhang2020new,rajpurkar2017mura}. To the best of our knowledge, no existing studies devoted to the development and evaluation of a comprehensive system for classifying and localizing multiple spine lesions from X-ray scans. The lack of large datasets with high-quality images and human experts' annotations is the key obstacle. To fill this gap, this work focuses on creating a significant benchmark dataset of spine X-rays that are manually annotated at the lesion level by experienced radiologists. We also propose to develop and evaluate, based on our dataset, a deep learning-based framework that includes a normal versus abnormal classifier followed by a lesion detector that localizes multiple categories of findings with bounding boxes. Both of these tasks can be beneficial in clinical practice: the normal versus abnormal classifier helps triage patients, while the lesion detector helps speed up the reading procedure and complements radiologist's observations.   

\subsection{Contribution}
Our contributions in this paper are two folds. First, we present a new large-scale dataset of 10,469 spine X-ray images from 5,000 studies that are manually annotated with 13 types of abnormalities by radiologists. This is the largest dataset to date that provides radiologist's bounding-box annotations for developing supervised-learning \emph{object detection} algorithms. Table~\ref{tab:data-set_overview} provides a summary of publicly available MSK datasets, in which two previous spine datasets~\cite{gertych2007bone,kim2020automatic} are significantly smaller than ours in size. Furthermore, human expert's annotations of spinal abnormalities are not available in those datasets. Second, we develop and evaluate VinDr-SpineXR -- a deep learning framework that is able to classify and localize multiple spine lesions. Our main goal is to provide a nontrivial baseline performance of state-of-the-art deep learning approaches on the released dataset, which could be useful for further model development and comparison of novel CAD tools. To facilitate new advances, we have made the dataset available for public access on our project's webpage\footnote{\url{https://vindr.ai/datasets/spinexr}}.  The codes used in the experiments are available at Github\footnote{\url{https://github.com/vinbigdata-medical/vindr-spinexr}}.
\begin{table}
\caption{Overview of publicly available MSK image datasets.
}
\label{tab:data-set_overview}
\centering
\begin{tabular}{|l|l|l|l|l|l|}
\hline
\textbf{Dataset}                                & \textbf{Year} &\textbf{Study type} & \textbf{Label} & \textbf{\# Im.} \\
\hline
Digital Hand Atlas~\cite{gertych2007bone}       & 2007 & Left hand          & Bone age              & 1,390 \\
\hline
Osteoarthritis Initiative~\cite{OAI}      & 2013 & Knee               & K\&L Grade            & 8,892 \\
\hline
MURA~\cite{rajpurkar2017mura}                   & 2017 & Upper body    & Abnormalities         & 40,561 \\
\hline
RSNA Pediatric Bone Age~\cite{halabi2019rsna}   & 2019 & Hand               & Bone age              & 14,236  \\
 \hline
Kuok \textit{et al.}~\cite{kuok2018vertebrae}   & 2018 & Spine              & Lumbar vertebrae mask & 60 \\
\hline
Kim \textit{et al.}~\cite{kim2020automatic}     & 2020 & Spine              & Spine position        & 797 \\
\hline
\textbf{Ours} & \textbf{2021} & \textbf{Spine} & \textbf{Multiple abnormalities} & \textbf{10,469} \\
\hline
\end{tabular}
\end{table}
\section{Proposed Method}
\label{sect:2}
\subsection{Overall framework} 
\begin{figure}
\includegraphics[width=12.5cm,height=5cm]{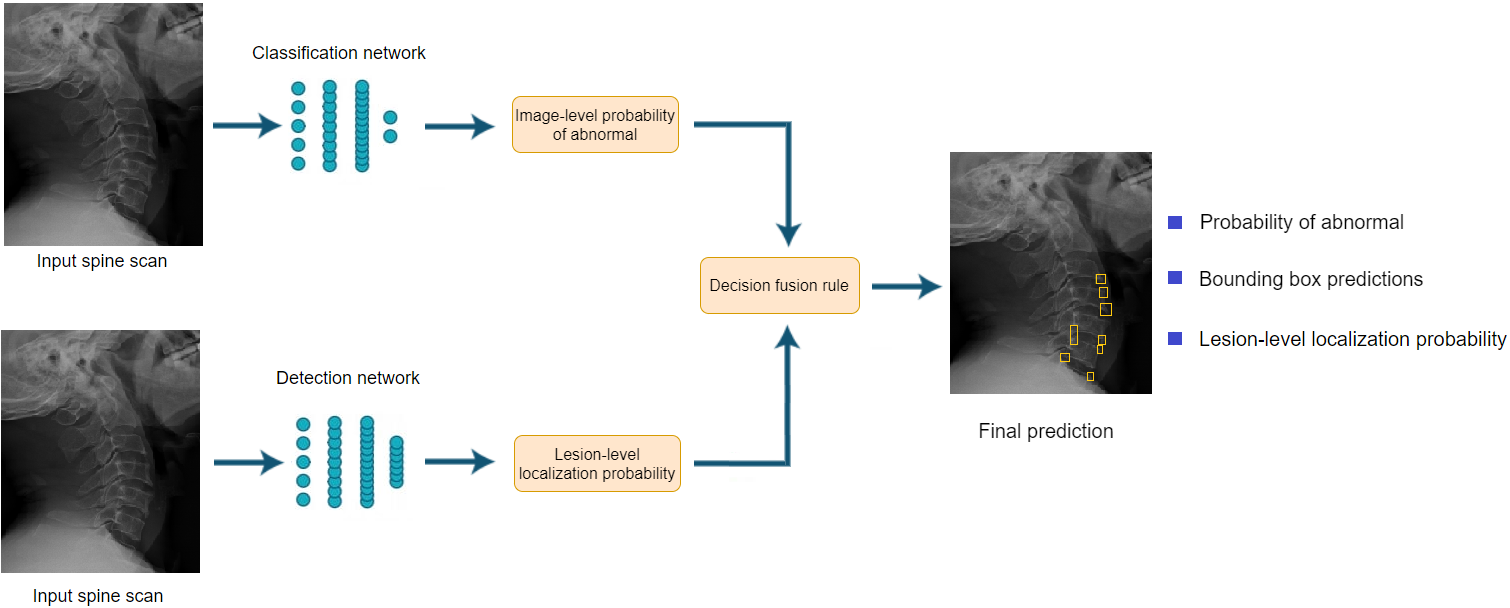}
\caption{Overview of the VinDr-SpineXR for spine abnormalities classification and localization. A binary classifier takes as input one spine scan and predicts its probability of abnormality. A detector takes the same scan as input and provides bounding boxes along with probabilities of abnormal findings at lesion-level. A decision rule is then proposed to combine the two outputs and maximize the detection performance.} \label{fig:proposed_system}
\end{figure}

 The proposed deep learning framework (see Figure~\ref{fig:proposed_system}) includes two main parts: (1) a classification network, which accepts a spine X-ray as input and predicts if it could be a normal or abnormal scan; (2) a detection network receives the same input and predicts the location of abnormal findings. To maximize the framework's detection performance, we propose a decision rule to combine the outputs of the two networks.

\subsection{Dataset}
\subsubsection{Data collection} Spine X-rays with lesion-level annotations are needed to develop automated lesion detection systems. In this work, more than 50,000 raw spine images in DICOM format were retrospectively collected from the Picture Archive and Communication System (PACS) of different primary hospitals between 2010--2020. The data collection process was conducted under our cooperation with participating hospitals. Since this research did not impact clinical care, patient consent was waived. To keep patient’s Protected Health Information (PHI) secure, all patient-identifiable information associated with the images has been removed. Several DICOM attributes that are important for evaluating the spine conditions like patient’s age and sex were retained.
\subsubsection{Data annotation} 
\begin{figure}
\includegraphics[width=12.5cm,height=2.5cm]{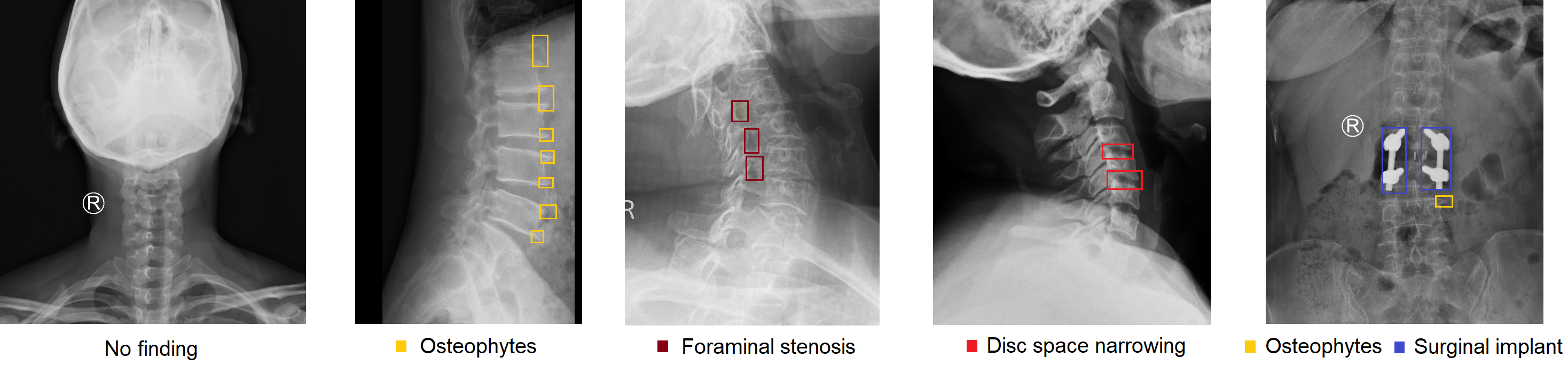}
\caption{Examples of spine X-rays with radiologist’s annotations, in which abnormal findings are marked by rectangular bounding boxes.} \label{fig:samples}
\end{figure}

To annotate data, we developed an in-house labeling framework called VinDr Lab~\cite{VinDrLab} which was built on top of a PACS. This is a web-based framework that allows multiple radiologists to work remotely at the same time. Additionally, it also provides a comprehensive set of annotation tools that help maximize the performance of human annotators. A set of 5,000 spine studies were randomly selected from the raw data after removing outliers (\textit{e.g.} scans of other body parts or low-quality). All these scans were then uploaded to the labeling framework and assigned to 3 participating radiologists, who have at least 10 years of experience, such that each scan is annotated by exactly one radiologist. In this step, the radiologists decide whether a scan was abnormal or normal as well as marked the exact location of each abnormality on the scan. The labels and annotations provided by the radiologists were served as ground truth for model development and validation later. During this process, the radiologists were blinded to relevant clinical information except patient's age and sex. Finally, a total of 10,468 spine images from 5,000 studies were annotated for the presence of 13 abnormal findings: ankylosis, disc space narrowing, enthesophytes, foraminal stenosis, fracture, osteophytes, sclerotic lesion, spondylolysthesis, subchondral sclerosis, surgical implant, vertebral collapse, foreign body, and other lesions. The ``no finding'' label was intended to represent the absence of all abnormalities. For the development of the deep learning algorithms, we randomly stratified the labeled dataset, at study-level, into a development set of 4,000 studies and a test set of 1,000 studies. Table~\ref{data-characs} summarizes the data characteristics, including patient demographic and the prevalence of each label via the number of bounding boxes. Figure~\ref{fig:samples} shows several representative samples with and without abnormal findings from our dataset.
\begin{table}
\centering
\caption{Characteristics of patients in the training and test datasets.}
\label{data-characs}
\begin{tabular}{p{10pt} | p{130pt}|p{55pt}|p{55pt} | p{55pt}}
\hline
& \textbf{Characteristic} & \textbf{Training set} & \textbf{Test set} & \textbf{Total} \\
\hline
  \parbox[t]{2mm}{\multirow{12}{*}{\rotatebox[origin=c]{90}{\textbf{Statistics}}}}
& Years                                 & 2011 to 2020       & 2015 to 2020       & 2011 to 2020 \\
& Number of studies                     & 4000               & 1000               & 5000 \\
& Number of images                      & 8390               & 2078               & 10468 \\
& Number of normal images               & 4257               & 1068               & 5325 \\
& Number of abnormal images             & 4133               & 1010               & 5143 \\ 
& Image size (pixel$\times$pixel, mean) & 2455 $\times$ 1606 & 2439 $\times$ 1587 & 2542 $\times$ 1602 \\
& Age (mean, years [range])*            & 49 [6 -- 94]       & 49 [13 -- 98]      & 49 [6 -- 98] \\
& Male (\%)*                            & 37.91              & 37.53              & 37.83 \\ 
& Female (\%)*                          & 62.08              & 62.47              & 62.17 \\
& Data size (GiB)                       & 29.04              & 7.17               & 36.21 \\ 
\hline
\parbox[t]{3mm}{\multirow{13}{*}{\rotatebox[origin=c]{90}{\textbf{Lesion type}}}}  
& 1.  Ankylosis             &     6 &    1 &     7 \\
& 2.  Disc space narrowing  &   924 &  231 &  1155 \\
& 3.  Enthesophytes         &    54 &   13 &    67 \\
& 4.  Foraminal stenosis    &   387 &   95 &   482 \\
& 5.  Fracture              &    10 &    2 &    12 \\
& 6.  Osteophytes           & 12562 & 3000 & 15562 \\
& 7.  Sclerotic lesion      &    24 &    5 &    29 \\
& 8.  Spondylolysthesis     &   280 &   69 &   349 \\
& 9.  Subchondral sclerosis &    87 &   23 &   110 \\
& 10. Surgical implant      &   429 &  107 &   536 \\
& 11. Vertebral collapse    &   268 &   69 &   337 \\
& 12. Foreign body          &    17 &    4 &    21 \\
& 13. Other lesions         &   248 &   63 &   311 \\
\hline
\end{tabular}
\begin{flushleft} 
(*) These calculations were performed on studies where sex and age were available.
\end{flushleft}
\end{table}
\subsection{Model development}
\subsubsection{Network architecture and training methodology} 
To classify a spine X-ray image as either normal or abnormal, three CNNs (DenseNet-121, DenseNet-169, DenseNet-201) have been deployed. These networks~\cite{huang2017densely} are well-known to be effective for X-ray interpretation~\cite{rajpurkar2017chexnet,rajpurkar2017mura}. Each network accepts an image of the spine as input and outputs a binary label corresponding to normal or abnormal scan. A total of 4,257 spine images with normal findings and 4,133 spine images with abnormal findings (reflecting any of the lesions) from the training set was used to optimize networks' weights. During this stage,  we optimized cross-entropy loss between the image-level labels and network's outputs using SGD optimizer. The average ensemble of three models serves as the final classifier. An image is considered abnormal in the inference stage if its probability of abnormality is greater than an optimal threshold. In particular, we determine  the optimal threshold $c^*$ for the classifier by maximizing Youden's index~\cite{youden1950index}, $J(c) = q(c) + r(c) - 1$, where the sensitivity $q$ and the specificity $r$ are functions of  the cutoff value  $c$. For the detection task, we aim to localize 7 important lesions: osteophytes, disc space narrowing, surgical implant, foraminal stenosis, spondylolysthesis, vertebral collapse, and other lesions. Due to the limited number of positive samples, the rest lesions were considered ``other lesions". State-of-the-art detectors, namely Faster R-CNN~\cite{ren2015faster}, RetinaNet~\cite{lin2017focal}, EfficientDet~\cite{tan2020efficientdet}, and Sparse R-CNN~\cite{peize2020sparse}, have been deployed for this task. Faster R-CNN~\cite{ren2015faster} was chosen as a representative for anchor-based two-stage detectors, which firstly propose a set of candidate regions then categorize and refine their locations. RetinaNet~\cite{lin2017focal} and EfficientDet~\cite{tan2020efficientdet} are both one-state detectors that directly localize and classify the densely proposed anchor boxes. Different from previous detectors, Sparse R-CNN~\cite{peize2020sparse} starts from a set of initial proposals, then repeatedly refines and classifies these boxes using attention mechanism. All the networks were trained to localize spine lesions using the stochastic gradient descent (SGD) optimizer. During the learning phase, the bounding box regression loss and region-level classification loss were jointly minimized. To improve the generalization performance of the detectors, learned data augmentation strategies for object detection~\cite{zoph2020learning} were incorporated in the training procedure.

\subsubsection{Decision fusion rule}  Given  an input image $\textbf{x}$, we denote $\hat{p}( \verb|abnormal| | \textbf{x})$ as the classifier's output that reflects the probability of the image being abnormal. To maximize the performance of the lesion detector, we propose a heuristic fusion rule as follows. For any $\textbf{x}$  with the prediction $\hat{p}( \verb|abnormal| | \textbf{x}) \geq c^*$, all lesion detection results are retained. For the case $\hat{p}( \verb|abnormal| | \textbf{x}) < c^*$, only predicted bounding boxes  with confidence higher than $0.5$ are kept. 
\section{Experiments and Results}
\label{sect:3}
\subsection{Experimental setup \& Implementation details}
Several experiments were conducted to evaluate the performance of the VinDr-SpineXR. First, we evaluated the classification and detection networks independently on the test set. We then investigated the effect of the fusion rule on the detection performance of the whole framework. All networks were implemented and trained using PyTorch framework (version 1.7.1) on a server with one NVIDIA V100 32GiB GPU. For the classification task, all training images were rescaled to 224 $\times$ 224 pixels and normalized by the mean and standard deviation of images from the ImageNet dataset. The networks were initialized with pre-trained weights on ImageNet. Each network was trained end-to-end using SGD for 10,000 iterations, approximately 2 hours. We used mini-batches of size 32 and set the learning rate to 0.001. An ensemble of the three best models served as the final classifier. For the detection task, all training images were randomly downsampled such that the shorter edge ranging between 640 and 800 pixels, then random augmentation transforms were applied before grouping into mini-batches of 16 samples. The detectors were initialized with weights pre-trained on COCO dataset then trained for 50,000 iterations (about 24 hours) using SGD with the learning rate reduced by $10\times$ at the 30,000-\textit{th} iteration. 
\subsection{Evaluation metrics}
We report the classification performance using AUROC, sensitivity, specificity, and F1 score. We also estimate the 95\% confidence interval (CI) by bootstrapping with 10,000 bootstraps for each measure. In each bootstrap, a new dataset is constructed by sampling with replacement~\cite{EfroTibs93}. For the detection task, we followed PASCAL VOC metric, mean average precision (mAP@0.5)~\cite{everingham2010pascal}. A predicted finding is a true positive if it has an intersection over union (IoU) of at least 0.5 with a ground truth lesion of the same class. For each class, average precision (AP) is the mean of 101 precision values, corresponding to recall values ranging from 0 to 1 with a step size of 0.01.  The final metric, mAP@0.5, is the mean of AP over all lesion categories.
\subsection{Experimental results}
\subsubsection{Classification performance}
On the test set of 2,078 images, our ensemble model reported an AUROC of 88.61\% (95\% CI 87.19\%, 90.02\%). At the optimal operating point of 0.2808 (see Supplementary Material), the network achieved a sensitivity of 83.07\% (95\% CI 80.64\%, 85.34\%), a specificity of 79.32\% (95\% CI 77.84\%, 81.71\%), and an F1 score of 81.06\% (95\% CI 79.20\%, 82.85\%), respectively. Table~\ref{tab:classification_result} shows quantitative results over all the classification networks.
\begin{table}
\centering
\caption{Classification performance on the test set (in percent with 95\% CI).}
\label{tab:classification_result}
\begin{tabular}{p{53pt}|p{67pt}|p{67pt}|p{67pt}|p{67pt} }
\hline
  \textbf{Classifier} & \textbf{AUROC} & \textbf{F1 score} & \textbf{Sensitivity} &  \textbf{Specificity} \\
\hline
    DenseNet121 & 86.93 (85.4,88.4)  & 79.55 (77.6,81.4) & 80.39 (77.9,82.8) & 79.32 (76.9,81.7) \\
    DenseNet169 & 87.29 (85.8,88.8)  & 80.25 (78.3,82.1) & 81.74 (79.3,84.1) & 79.02 (76.6,81.4) \\
    DenseNet201 & 87.14 (85.6,88.6)  & 79.03 (77.1,80.9) & 77.97 (75.4,80.5) & 81.46 (79.1,83.8) \\
    \hline
    Ensemble    & 88.61 (87.2,90.0)  & 81.06 (79.2,82.9) & 83.07 (80.6,85.3) & 79.32 (77.8,81.7) \\
\hline
\end{tabular}
\end{table}
\subsubsection{Detection performance}
The AP of each abnormal finding detected by 4 detectors is shown in Table~\ref{tab:detection_result}. Sparse R-CNN~\cite{peize2020sparse} showed it as the best performing detector for this task with a mAP@0.5 of 33.15\% (see Supplementary Material). Meanwhile, RetinaNet~\cite{lin2017focal} showed the worst performance with a mAP@0.5 of 28.09\%. We observed that the reported performances varied over the target lesions, \textit{e.g.} all the detectors performed best on the LT10 label (surgical implant) and worst on the LT13 (other lesions).
\begin{table}
\centering
\caption{Spine X-ray detection performance on the test set.}
\label{tab:detection_result}
\begin{tabular}{p{78pt}|p{27pt}|p{25pt}|p{25pt}|p{25pt}|p{27pt}|p{27pt}|p{27pt}|p{45pt}}
\hline
 \textbf{Detector} & \textbf{LT2$^{(*)}$} & \textbf{LT4} & \textbf{LT6} & \textbf{LT8} & \textbf{LT10} & \textbf{LT11} & \textbf{LT13} & \textbf{mAP@0.5} \\
\hline
    
    Faster R-CNN~\cite{ren2015faster}        & 22.66 & 35.99 & 49.24 & 31.68 & 65.22 & 51.68 & 2.16 & 31.83 \\
    RetinaNet~\cite{lin2017focal}            & 14.53 & 25.35 & 41.67 & 32.14 & 65.49 & 51.85 & 5.30 & 28.09 \\
    EfficientDet~\cite{tan2020efficientdet}  & 17.05 & 24.19 & 42.69 & 35.18 & 61.85 & 52.53 & 2.45 & 28.73 \\
    Sparse R-CNN~\cite{peize2020sparse}      & 20.09 & 32.67 & 48.16 & 45.32 & 72.20 & 49.30 & 5.41 & 33.15 \\
\hline
\end{tabular}
\begin{flushleft} 
(*) LT2, LT4, LT6, LT8, LT10, LT11, LT13 denotes for disc space narrowing, foraminal stenosis, osteophytes, spondylolysthesis, surgical implant, vertebral collapse and other lesions, respectively, following the same indexing in Table \ref{data-characs}.
\end{flushleft}
\end{table}

\subsubsection{Effect of the decision fusion rule} Table~\ref{tab:compare_framework} provides a comparison of the detection performance between the detector network and the whole framework. We observed that by combining the detection and classification networks via the proposed decision rule, the whole framework leads to a higher performance over all lesions, except LT4 and LT8. Visualization of predicted lesions by the VinDr-SpineXR is provided in the Supplementary Material.
\begin{table}
\centering
\caption{Spine X-ray detection performance of the whole framework on the test set.}
\label{tab:compare_framework}
\begin{tabular}{p{78pt} |p{23pt}|p{23pt}|p{23pt}|p{23pt}|p{23pt}|p{23pt}|p{23pt}|p{45pt}}
\hline
 \textbf{Method} & \textbf{LT2} & \textbf{LT4} & \textbf{LT6} & \textbf{LT8} & \textbf{LT10} & \textbf{LT11} & \textbf{LT13} & \textbf{mAP@0.5} \\
\hline
  
  Detector only     & 18.06 & 28.89 & 34.23 & 41.52 & 62.45 & 42.85 & 4.03 & 33.15 \\
  Whole framework   & 21.43 & 27.36 & 34.78 & 41.29 & 62.53 & 43.39 & 4.16 & 33.56 \\
\hline
\end{tabular}
\end{table}

\section{Discussion}
The proposed decision fusion uses the result of the classifier to influence the detector. As shown in Table \ref{tab:compare_framework}, this rule, although very simple, helps improve the mAP of the detector by \textbf{0.41\%}. We have also experimented with a counterpart fusion rule to use the result of the detector to influence the classifier. Specifically, we averaged classifier's output with the highest box score from detector's outputs, boosting the AUROC and F1 score of the classifier by \textbf{1.58\%} and \textbf{0.46\%}, respectively. These experiments have highlighted the effectiveness of the proposed mutual ensemble mechanism.
\section{Conclusions}
\label{sect:4}
In this work, we developed and validated VinDr-SpineXR -- a deep learning-based framework to detect and localize abnormalities in radiographs of spine. We contributed a new labeled dataset and models for spine X-ray analysis. To the best of our knowledge, this is the first effort to address the problem of multiple lesions detection and localization in spine X-rays. Our experiments on the dataset demonstrated the effectiveness of the proposed method. For future work, we expect to extend our dataset and consider more lesion labels for further experiments, including rare labels. We also plan to conduct more experiments and evaluate the impact of the proposed framework in real-world clinical scenarios.
\bibliographystyle{splncs04}
\bibliography{refs}
\end{document}